\documentstyle[12pt]{article}

\catcode`@=11  \@addtoreset{equation}{section}  \catcode`@=12
\def\be{\begin{equation}}
\def\ee{\end{equation}}
\def\IP{\hbox{\rm I\kern -1.6pt{\rm P}}}
\def\IC{{\hbox{\rm C\kern-.58em{\raise.53ex\hbox{$\scriptscriptstyle|$}}
    \kern-.55em{\raise.53ex\hbox{$\scriptscriptstyle|$}} }}}
\def\IN{\hbox{I\kern-.2em\hbox{N}}}
\def\IR{\hbox{\rm I\kern-.2em\hbox{\rm R}}}
\def\IZ{\hbox{{\rm Z}\kern-.3em{\rm Z}}}
\def\IT{\hbox{\rm T\kern-.38em{\raise.415ex\hbox{$\scriptstyle|$}} }}
\def\notsub{\hbox{$\subset$\kern-.55em\hbox{/}}}

                    \def\mod1{\,({\rm mod\ } 1)\,}
\def\ep{\varepsilon}   \def\phi{\varphi}    
\def\1#1{\hbox{\large \bf 1}_{#1}}     % unit function on #1
\def\I|#1{{\big|}_{#1}}                % restriction to #1

\def\epd#1{\if #11 \ep^{-d} \else \ep^{-#1d} \fi}  % \epsilon^{-#1d}
\renewcommand{\topmargin}{-38pt}

\begin{document}
\title{ The decay of multiscale signals --
deterministic model of the Burgers turbulence }
\author{S.N. Gurbatov$^{1,3}$, A.V.Troussov$^{2,3}$}
\maketitle
\begin{tabular}{ll}
$^1$ &Radiophysics Dept., University of Nizhny Novgorod\\
& 23, Gagarin Ave., Nizhny Novgorod 603600, Russia. \\
& E-mail: gurb@rf.unn.runnet.ru\\
&(Permanent address) \\

$^2$& Joint Institute of Physics of the Earth RAS,  \\
& Molodezhnaya Str., 3, GC RAS, Moscow, 117296
Russia. \\
& E-mail: shtrssv@cityline.ru, troussov@hotmail.com\\
&(Permanent address) \\
$^3$ &Observatoire de la Cote d'Azur, Laboratorie G.D. Cassini,\\
& B.P. 229, 06304, Nice Cedex 4, France.\\
\end{tabular}

\date{}

\noindent{\bf Corresponding author:}
        %Alexandre V. Troussov, JIPE RAS,
        %Molodezhnaya Str.3, GC RAS, Moscow, 117296
        %Russia. E-mail: troussov@hotmail.com
        %Tel.: (7-095) 1334339, Fax: (7-095) 9305559, Tel.(Home): (7-095)
3731007.
Sergey N. Gurbatov, Radiophysics Dept.,\\
 University of Nizhny Novgorod,
23, Gagarin Ave., Nizhny Novgorod 603600, Russia.\\
E-mail: gurb@rf.unn.runnet.ru.\\
Tel.: (8 312) 656002, Fax: (8 312) 656416,\\ Tel.(Home): (8
312) 340969.

\begin{abstract}

This work is devoted to the study of the decay
of multiscale deterministic solutions of the
unforced Burgers' equation in the limit of vanishing viscosity.

It is well known that Burgers turbulence with
a power law energy spectrum $E_0(k) \sim {|k|}^n$ has
a self-similar regime of evolution.
For $n<1$ this regime is characterised by an integral scale
$L(t) \sim t^{2/{(3+n)}}$, which increases with the time due to the
multiple mergings of the shocks, and therefore,
the energy of a random wave decays more slowly
than the energy of a periodic signal.

In this paper a deterministic model of turbulence-like
evolution is considered. We construct the initial perturbation
as a piecewise linear analog of the Weierstrass function.
The wavenumbers of this function form a "Weierstrass spectrum",
which accumulates at the origin in  geometric progression.
"Reverse" sawtooth functions with negative initial slope
are used in this series as basic functions, while their
amplitudes   are chosen by the condition that
the distribution of energy over exponential intervals of wavenumbers
is the same as for the continuous spectrum in Burgers
turbulence. Combining these two ideas allows us to obtain
an exact analytical solution for the velocity field.
We also notice  that such multiscale waves may be constructed
for multidimensional Burgers' equation.

This solution has scaling exponent $h=-(1+n)/2$
and its evolution in time is self-similar with logarithmic
periodicity and with the same average law $L(t)$ as for
Burgers turbulence.  Shocklines form self-similar regular
tree-like structures. This model also describes important properties
of the Burgers turbulence such as the self-preservation of
the evolution of
large scale structures in the presence of small scales perturbations.

\end{abstract}



%=======================================================================
\section{Introduction}\label{Introduction} \vskip 4mm

The nonlinear diffusion equation
\begin{equation}
   { \partial v \over \partial t   } + v {\partial v\over \partial x}
      = \nu {\partial^2 v \over\partial x^2 } ;\ \  v(x,t=0) = v_0(x) .
                                         \label{eq:DifEq}
\end{equation}
was originally introduced by J.M.Burgers in \cite{Burgers39} (1939)
as a model for hydrodynamical turbulence.
Burgers' equation (\ref{eq:DifEq})
describes two fundamental effects characteristic of any turbulence
\cite{FrischTurbulence95}:
the nonlinear redistribution of energy over the spectrum and
the action of viscosity in small scales. Burgers' equation was used later to
describe a large class of physical systems in which the nonlinearity
is fairly weak (quadratic) and the dispersion is negligible
compared to the linear damping \cite{Whitham}.
The most important example of such waves are
acoustical waves with finite amplitude \cite{RudenkoSoluyan}.
Another class of problems, arising, e.g.,
in surface growth, also leads to Burgers' equation
\cite{KPZ},\cite{BarSt},\cite{WW98}. The three dimensional form
of (\ref{eq:DifEq})
has been used in cosmology to describe the
formation of large scale structures
of the Universe at a nonlinear stage of gravitational instability
(~see e.g.~\cite{GSS89},~\cite{SZ89},~\cite{GMS91},~\cite{VeDuFrNo94}~).

In the physically important case of large Reynolds number, the action
of viscosity is significant only in the small regions with
high gradient of the velocity field.
In the limit \  $\nu\rightarrow 0$,
the solution of Burgers' equation has the following form
(see \cite{Hopf50}, \cite{Burgers74}, \cite{GMS91}):
\begin{equation}
     v(x,t) =  { x - y(x,t) \over t },  \label{eq:burgers}
\end{equation}
where $y(x,t)$ is the coordinate of the maximum of the function
\begin{equation}
     G(x,y,t) = \Psi_0 (y) - { (x-y)^2 \over 2t }, \
     v_0(x) = -{ {\partial \Psi_0 (x)} \over {\partial x} }.
  \label{eq:?}
\end{equation}

Strong interaction between coherent harmonics leads
to the appearance of local self-similar structures in Burgers' equation.
A periodic initial perturbation with zero mean velocity
is transformed asymptotically into a sawtooth wave with gradient \
$\partial_x v = { 1 / t} $ \ and with the same period $l_0$.
It is important that at this stage the amplitude
$a(t)={l_0 / t}$ and the energy density
$ \sigma^2(t) \simeq {l_0^2 / 12 t^2}$
do not depend on the initial amplitude.

An initial one-signed pulse with the area $m > 0$, localised at
$t=0$ in the neighbourhood of the point \ $x=0$,\  also has asymptotically
a universal form: it transforms into a triangular pulse with the gradient
$ \partial_x v = 1 / t $
and increasing coordinate of the shock $ x_s \approx (2mt)^{1/2}$.
Due to the increase of the integral scale the amplitude of such a pulse
$a(t) = x_s(t) / t ~  \sim m^{1/2}t^{-1/2} $\
and its energy
will decrease more slowly  than for a periodic signal, like
$t^{-{1/2}}$.

Continuous random initial fields are also transformed into
sequences of regions with the same gradient $\partial_x v = 1/t$,
but with random locations of the shocks separating them.
Due to the multiple merging of the shocks
the statistical properties of such random fields
are also self-similar and may be characterised by the
integral scale of the turbulence $L(t)$.
The merging of the shocks
leads to an increase of the integral scale $L(t)$, and because of this
the energy
\begin{equation}
  \sigma^2(t) \sim L^2(t) / t^2  \label{eq:sigmat} % \ref{eq:sigmat}
\end{equation}
of a random wave decreases more slowly than the energy of periodic signals.

The type of turbulence evolution is determined by the behaviour
of the large scale part of the initial energy
spectrum
\begin{equation}
   E_0(k) = \alpha^2 k^n b_0(k);   \label{eq:no4}
\end{equation}
\begin{equation}
   E_0(k) = { 1 \over 2 \pi } \int \langle v_0(x),v_0(x+z) \rangle e^{ikz}
dz.
\end{equation}
Here \ $b_0(k)$\  is a function which falls off rapidly
for \ $k > k_0 \sim l_0$, and \ $b_0(0) = 1$. For $n>1$  the
law of enrgy decay strongly depends on the statistical
properties of the initial field (see e.g.
\cite{WW98} and references therein).
For the initial Gaussian perturbation
the integral scale $L(t) \sim t^{1/2}$
times logarithmic correction obtains and is determined
by two integral characteristics of the initial
spectrum:
the variances of the initial potential $\Psi_0$
and the velocity $v_0(x)$~
\cite{Kida79},\cite{FournierFrisch83},\cite{GMS91},\cite{GSAFT97}.

For $n < 1$  the structure function of the initial potential
increases as a power
law in space.  Then the initial potential field is Brownian, or
fractional Brownian  motion, and some scaling may be used
\cite{Burgers74},\cite{Kida79}, \cite{GMS91}, \cite{sinai92},
\cite{AvellanedaE95},\cite{Molchan97}, \cite{Ryen98}.
In this case the  turbulence is also self-similar and
the integral scale $L(t)$ increases as
\begin{equation}
   L(t) = (\alpha t)^{ 2 / (3+n)}.   \label{eq:lt}
\end{equation}
The energy of the turbulence is derived from (\ref{eq:sigmat}):
\begin{equation}
 \sigma^2 (t) \sim t^{-p}, \ p = {{2(n+1})\over{n+3} }\ . \label{eq:sigmap}
\end{equation}

The difference between these two cases  \ ( $n<1$ and $n>1$ )
is connected to the process of parametric generation of low
frequency component of the spectrum. For the case $n<1$
the newly generated low frequency components are relatively small
and we have the conservation of large scale part of the spectrum:
\begin{equation}
   E(k,t) = E_0(k) = \alpha^2k^n, \ \ for\  k<< 1/ L(t). \label{eq:energy}
\end{equation}

Thus, the laws of turbulent decay are more complex than
for simple signals, which can be attributed to
multiple merging of the shocks.
In \cite{GuCr95} a model of a regular fractal signal with decay
lower than for single one-signed pulse was introduced.
The initial signal $v_0(x)$ was constructed
as a sequence of one-signed pulses whose positions form a Cantor set
with capacity (fractal dimension) $D=\ln N / \ln \beta$,
where $N^p$ is the number of pulses in the scale
$ L_p \simeq L_1 N \beta^{p-1}$,\  $0<D<1$.
Multiple merging makes the decay of the wave slower
and the general behaviour of the energy decay may be
approximated by the power law with the exponent
in (\ref{eq:sigmap}):
\begin{equation}
   p = {{1-D} \over {2-D}}, \ \ \   0<p<1.   \label{eq:pf}
\end{equation}
The evolution proves to be self-similar in successive
time periods $ (t_i,t_{i-1})$ and $ (t_{i+1},t_{i+2})$,
where $ t_{i+1} / t_i = \beta^2 / N $.
This shows log-periodical self-similarity of the field evolution.
Linear and non-linear decay of fractal and spiral fields
given by the sequences of regular pulses
was also investigated in \cite{AngilellaVassilicos98}.
It was shown that the power law  (\ref{eq:sigmap})
with the exponent given by the formula (\ref{eq:pf})
holds also true for homogeneous fractal pulse signal with capacity $D$.

Another model of a multiscale signal, which has the same
general behaviour on the external scale $L(t)$ (\ref{eq:lt}), and
energy of the Burgers turbulence, was also discussed in \cite{GuCr95}.
It was assumed therein that the initial signal is
a discrete set of modes - the spatial harmonics
\begin{equation}
   v_0(x) = \sum_{p=0}^{\infty} a_p \sin (k_p x + \phi_p), \label{eq:1.7}
\end{equation}
with wavenumbers $k_p$ and amplitudes $a_p$
given in terms of a parameter $\epsilon$ by
\begin{eqnarray}
 k_p & = & k_0 \epsilon^p, \ \ \ \ \ \ \ \ \  a_p = a_0 \epsilon^{-hp},
 \nonumber \\
 h   & = & -( { {n +1} \over 2} ), \ \  a_0 = \alpha k_0^{(n+1)/2}.
                                      \label{kpapa0}
 \end{eqnarray}
Amplitudes $a_p$ and the scaling exponent $h$ are chosen
from the condition that the mean energy of harmonics in the
interval $\Delta_p = k_p - k_{p+1}$ be identical to that
corresponding to the spectral density (\ref{eq:no4}):\ $a_p^2 =
E(k_p) \Delta_p$. \ For $\epsilon << 1$ and $ n > -1$ the harmonics
are spread over the spatial spectrum and accumulate
at the point $ k=0$ with decreasing amplitude.
The main approach in this model was that the energy of the wave
is the sum of energies of independent modes.
The approach is nontrivial, but nevertheless
leads to the same laws of the integral scale $L(t)$ (\ref{eq:lt})
and the energy decay (\ref{eq:sigmap})
as in the case of continuous spectrum (\ref{eq:no4}).
Let us point out that representation
of the field given by the formula (\ref{eq:1.7}) is similar to shell
models, which were introduced as useful models addressing the
problem of analogous scaling in fully developed turbulence (see,
e.g., \cite{GilsonDaumondDombre98}, \cite{Kadanoff98} and references
in there).

In present paper we consider the evolution
of a regular signal whose behaviour in general
is similar to the evolution of the Burgers turbulence
with continuous spectrum (\ref{eq:no4}).
The main difference,as compared to
the model discussed above, is that we
construct the exact solution of Burgers' equation using as an initial
mode the "reverse" sawtooth wave.
The frequency ratio in our model is $\epsilon=1/N$,
where $N$ is an integer and $N \geq 2$.
These perturbation are similar to the well known
Weierstrass and Weierstrass-Mandelbrot fractal functions
(see \cite{Mandelbrot77}, \cite{BerryLewis1980}).

For the analysis of Burgers' equation it
is convenient to use a mechanical interpretation.
There is a one-to-one correspondence between the solution
of Burgers' equation and the dynamic of a gas of inelastically
interacting particles \cite{GMS91}, \cite{GuSa93}.
Let us take a one-dimensional particle flux
with a contact interaction: as long as the particles do not
run into each other they move with  constant
velocity. In the collision they stick together,
forming a delta-function singularity in the matter
density. This leads to the appearance of gas of
two species: a hydrodynamical flux of the
"light" initial particles, and a gas of "heavy"
particles arising in the adhesion process of light particles.
The evolution of the particle velocity field
will be described by the solution of Burgers' equation
if we assume that the initial density of the light particles
is $ \rho_0 = const$,
the velocity of particles is equal to the
initial velocity in (\ref{eq:DifEq}), and that the
collision of the particles conserve their mass
and momentum. This analogy permits construction of a
very fast (linear time) algorithm of solution
of Burgers' equation \cite{TroussovLinAlg96}.

In our case, for the initial reverse sawtooth wave,
all the matter turns into heavy particles
at the same moment of time.
Thus, after this time the evolution of the Burgers turbulence is
fully determined by the motion of  heavy
particles, whose positions are  positions of
shocks, and
 masses are equal to $ \Delta v \cdot t $, where
$\Delta v$ are the amplitudes of the shocks.

The paper is organized as follows. In Section 2  we consider
the evolution and interaction of ``reverse'' sawtooth modes.
In section 3 we consider the interaction of  small scale mode with
large scale structures. In
Section 4 we investigate the properties of the sawtooth
Weierstrass-Mandelbrot fractal function. In section 5 we show that
deterministic model has logarithmic periodic self-similarity.
We also discuss  here the multi-dimensional generalization of this model.
Section 5 presents concluding remarks.
%=======================================================================
\section{Evolution and interaction  of "reverse" sawtooth modes in
Burgers' equation }\label{} \vskip 4mm

Let us introduce the $p$th "reverse"
saw tooth mode as
\begin{equation}
    v_{(p)}(x,0) = a_p A ( k_p x + \phi_p) \label{eq:2.1}
\end{equation}
Here $a_p, k_p$ are amplitude and wavenumber of the mode,
$-\phi_p$ is its phase.
The function $A(x)$ is $2\pi$ periodic function,
given on its first period by the following expression
\begin{equation}
    A(x) = \pi - x, \ \ \ x \in [0,2 \pi[ \label{eq:2.2}
\end{equation}
The set of "reverse sawtooth" functions is not orthogonal, but
nevertheless we will introduce a set of modes
satisfying equation (\ref{kpapa0}),
whose wavenumbers and amplitudes satisfy the same relations
as the sinusoidal modes
of \cite{GuCr95}, i.e. relation (\ref{eq:1.7}).
We introduce the term "reverse sawtooth"
because this signal is
a sawtooth with teeth facing to the right,
but the term "sawtooth" by itself is widely used
in Burgers turbulence literature
to refer to the late stage of the evolution
of the wave profile a sequence
of sawteeth with {\em positive} slope $1/t$.

The solution of Burgers' equation with a linear velocity profile
$ v_0 = - \gamma ( x - x_+) $
is well-known (see, e.g., \cite{GMS91}):
\begin{equation}
    v(x,t) = { - \gamma ( x - x_+) \over 1 - \gamma t } \label{eq:2.4}
\end{equation}
The value $\gamma^{-1}$ has the dimension of time, and for
$\gamma > 0 $, at the finite time $t = \gamma^{-1}$ the gradient
\ $ \partial_x v$ \  becames infinite. For $\gamma < 0 $ the gradient
becomes equal to $ \partial_x v =  t^{-1}$, independent of the value
of $\gamma$ at times $t \gg | \gamma|^{-1}$.
Thus, we have from (\ref{kpapa0}), (\ref{eq:2.1}),(\ref{eq:2.2}),
that the evolution of the $p$th mode is characterised by the nonlinear
time
\begin{equation}
    t_p = \gamma_p^{-1} = {a_pk_p}^{-1} = t_0/( \epsilon^{(n+3)/2})^p,
    \ \ t_0 = 1 / \alpha k_0^{n+3/2} \label{eq:2.5}
\end{equation}
Based on solution (\ref{eq:2.4} ) it is easy to see that for the
"first" period (if $\phi_p=0$) the evolution of the $p$th reverse mode
at the initial stage $(t< t_p)$ may be described as
\begin{equation}
v_{(p)}(x,t) = \cases{ x/t, &if $0< x < {t\over
t_p}{\pi \over k_p}$; \cr {1\over t_p}({\pi\over k_p}-x) / {1 \over
{1-t/t_p}}, & if $| { \pi\over k_p}-x | < {t\over t_p}{\pi\over k_p}$
       ;    \cr ({ x-{{\pi}\over  k_p}}) / t, & if
       ${t\over t_p}{\pi \over k_p}
             < x - {\pi\over k_p} < {\pi\over k_p} $ .  }   \label{eq:2.6}
\end{equation}
On the other hand, at time $t = t_p$,
the mode transforms into a "direct" sawtooth wave
with slope $\partial_x  v = {1/t}$\ ,\
independent of the amplitude and wavenumber of the mode:
\begin{equation}
 v_{(p)}(x,t) = \cases{ x/t, &if $0< x <
     {\pi \over k_p}$; \cr { (x - {\pi \over
      k_p}}) / t, & if $ { \pi\over k_p} < x  < {2\pi\over
      k_p}$  . \cr }   \label{eq:2.7}
\end{equation}
The density of energy $\sigma^2(t) = \langle v^2(x,t) \rangle_L$, where
$\langle \rangle_L$ denotes averaging over the period,
is conserved before $t<t_p$,
and decreases like $(k_pt)^{-2}$ after $t>t_p$.

Consider now the evolution of a gas of sticky particles
in the case of independent evolution of the $p$th mode.
In the general case, the density of the gas is calculated by using
the Jacobian of the transformation from Lagrangian to
Eulerian coordinates and may be written in the form
(~see, e.g., \cite{GMS91}~)
\begin{equation}
   \rho(x,t) = \rho_0 (1-t\partial_x v(x,t))   \label{eq:rhx}
\end{equation}
Then it is obvious that at the initial stage
\begin{equation}
 \rho(x,t) = \rho_0 { 1\over { 1 - {t/t_p}}
    }, \ \ \ \  |{\pi\over k_p}-x| < {t\over t_p} {\pi\over k_p}
     \label{eq:2.7a}
\end{equation}
while $\rho$ is zero outside this interval in each period.
At time $t=t_p$ all the light
particles in each period collide into a single heavy particle with mass
\begin{equation}
   m_p = \rho_0 L_p = \rho_0 {2\pi\over k_p} ,
     \label{eq:2.8}
\end{equation}
and the heavy particles have positions
\begin{equation}
 x_{p,l} = {\pi\over k_p} - {\phi_p\over k_p} + {2\pi\over k_p}l; \ \ \
                    l = 0,\pm 1, \pm2, ...       \label{eq:2.9}
\end{equation}
equal to the zero positions of the initial $p$th mode.
The process of light merging particles  and the evolution
of the velocity is shown in Fig.1.

Consider now the joint evolution of two successive
modes: $p$th and $(p+1)$th. From (\ref{eq:2.5}) one can see
that the ratio of nonlinear times of the successive modes is
\begin{equation}
  { t_{p+1} \over t_p } = \epsilon^{ - (n+3)/2 } \equiv \epsilon^{1-h},
     \label{eq:2.19}
\end{equation}
which does not depend on $p$ and increases
if the exponent $n$ is greater than $-3$.
The gradient of the initial field \ \  $v_{(p)}(x) + v_{(p+1)}(x) $
\ \ is \ \ \ $-(\gamma_p+\gamma_{p+1})$, \  so the effective
nonlinear time for such a sum is
\begin{equation}
   t_{p,eff} = t_{p,p+1} = {1 \over {\gamma_p + \gamma_{p+1}}} = {
   t_p \over {1+t_p/t_{p+1}} }, \label{eq:2.20}
\end{equation}
Because all parts of the initial perturbation have the same slope,
all light particles will collide at the same time $t = t_{p,p+1}$.

The mass of heavy particles after merging
are $m_{p,i} = \rho_0 \Delta_{i,i+1}$, where
$\Delta_{i,i+1}$ is the distance between adjacent shocks
in the initial perturbation. The ratio of periods of two adjacent modes
is $L_{p+1}/L_p = k_p / k_{p+1} =  \epsilon^{-1} $.
If $N=\epsilon^{-1}$ is an integer larger than $1$,
there will be $N+1$ heavy particles
on the period of the larger scale mode $(p+1)$th:
\ \  $(N-1)$ with the mass
$m_p = \rho_0 L_p$ \ (\ref{eq:2.8}), and two particles with total
mass equal to $m_p$. These two particles only exist when the shock
of the $(p+1)$th mode is located in the interval between shocks
of the $p$th mode. For simplicity, we will consider the case where
the spatial relations
\begin{equation}
   k_{p+1} \phi_{p+1} = k_p \phi_p + 2\pi {r/N},  \label{eq:2.21}
\end{equation}
between the phases of successive modes, hold. In this case the
discontinuities of the $(p+1)$th mode do not produce new shocks in
the total perturbation $v_{(p)}(x) + v_{(p+1)}(x)$. Thus, at times
$t$ larger than $t>t_{r,eff}$, the masses of all heavy particles
will be the same, as would be the case without the large scale
modes (\ref{eq:2.8}).

The positions of these heavy particles at time $t=t_{p,eff}$ are
\begin{equation}
   X_{(p,l)} (t_{p,eff}) = x_{p,l} + v_{(p+1)}(x_{p,l})
   t_{p,eff} ,  \label{eq:2.22}
\end{equation}
where $x_{p,l}$  are the zero positions of the $p$th mode (\ref{eq:2.9}).
The velocity of this particle is equal  to $v_{(p+1)}(x_{p,l})$.
Equation (\ref{eq:2.22}) is obvious if we use the trivial
equality $v_{(p)}(x_{p,l})+v_{(p+1)}(x_{p,l})=v_{(p+1)}(x_{p,l})$,
and also note that the position of the heavy particle
$x_{p,l}(t_{p,eff})$ is equal to the position at the same time
of all light particles with initial coordinate $x=x_{p,l}$. From
(\ref{eq:2.22}) we immediately have that after time $t_{p,eff}$
the positions of the particles are
\begin{equation}
   X_{(p,l)} (t) = x_{(p,l)} + v_{(p+1)}(x_{p,l}) t . \label{eq:2.23}
\end{equation}
The difference between the coordinates of the adjacent particles
$X_{p,l}(t)$ and $X_{p,l+1}(t)$ decreases with time, proportionally
to the gradient of $v_{(p+1)}(x)$:
\begin{eqnarray}
X_{(p,l+1)}(t) - X_{p,l}(t) =
(x_{p,l+1}-x_{p,l})-t{\partial v_{(p+1)}(x)\over\partial
x}(x_{p,l+1}-x_{p,l})
       \equiv   \nonumber \\
      (x_{p,l+1} - x_{p,l} ) (1 - {t / t_{p+1}}). \ \ \ \ \ \ \
     \label{eq:2.24}
\end{eqnarray}

These particles collide at time $t=t_{p+1}$ (\ref{eq:2.5})
and the newly created heavy particles will have masses
\begin{equation}
   m_{(p+1)} = \rho_0 L_{p+1} \label{eq:2.25}
\end{equation}
and positions
\begin{equation}
   x_{p+1,l} =
   { \pi\over k_{p+1} } - {\phi_{p+1}\over k_p} + {2\pi l\over k_{p+1}};
    \ \ \ l = 0,\pm 1, \pm2, ...
     \label{eq:2.26}
\end{equation}
The velocity of these particles is zero.

Thus, at times $t$  larger then $t_{p+1}$, the evolution of the initial
perturbation $ v_0(x) = v_{(p)}(x) + v_{(p+1)}(x)$ will be the same
as the evolution of only the large scale mode $v_{(p+1)}(x)$.
The process of particles merging and the evolution of the velocity
for the sum of two successive modes with the periods ratio
$N=2$ are shown in  Fig. 2.

By recurrence, it is evident that for finite number
of modes $v(x) = v_{(p)}(x) + v_{(p+1)} (x) + ... + v_{(M)}(x)$
the evolution of the field after $t_M$ will
be the same as the evolution of only the largest mode $v_{(M)}(x)$.
The reason for this, is of course, the special relation between
the phases $\phi_p$ and
the wavenumbers $k_p$ of all interacting modes: $ k_p = k_0 / N^p$
(~see equation (\ref{kpapa0})~).
For integer $N$ the minimal value of
any combination of these wavenumbers is equal to
the largest mode wavenumber $ k_M = k_0 / N^M$.
So the nonlinear interaction does not produce
new components at frequencies less than $k_M$.
%=======================================================================
\section{Interaction of small scale "reverse"  sawtooth mode
with large scale structures }\label{} \vskip 4mm

Let us now consider the interaction of the $p$th mode
with an infinite series of larger scale modes
\begin{equation}
   W_p(x)=\sum_{r=p+1}^{\infty}v_r(x)
   \equiv\sum_{r=p+1}^{\infty}a_r A(k_r x + \phi_r),   \label{eq:2.27}
\end{equation}
assuming that the phases of the modes satisfy the relations
(~\ref{eq:2.21}~)
and that
\ $k_r=k_0 \epsilon^r,\  a_r=a_0 \epsilon^{-hr},\
h = - (n+1) / 2 $. From (\ref{eq:2.27})
and (\ref{eq:2.5}) we have for the gradient of the initial
perturbation $ v_0(x) = v_p(x) + W_p(x) $
\begin{eqnarray}
\partial_x v_0(x) &= \partial_xv_p(x)+\partial_x W_p(x) =
                      \sum_{r=p}^\infty \gamma_p =     \nonumber \\
   &= \gamma_0 \sum_{r=p}^\infty ( \epsilon^{{(n+3)}\over 2})^r =
   \gamma_0 \epsilon^{{(n+3)\over 2}p}
   {1 \over { 1 - { \epsilon^{(n+3) \over 2}} }},
   \label{eq:2.28}
\end{eqnarray}
the condition $n >-3$ ($h<1$) being necessary for the
series to converge. From (\ref{eq:2.28}), we
have for the effective time of nonlinearity of $p$th mode
\begin{equation}
   \mathaccent126t_p = {1 / {\partial_x v_0(x)}}
   = t_p ( 1 - { \epsilon^{(n+3) \over 2}} ),      \label{eq:2.29}
\end{equation}
with the original $t_p$ determined by the equation (\ref{eq:2.5}).

Thus, after the time of collision $\mathaccent126t_p$, heavy particles
with mass $ m_p = \rho_0 L_p$ (\ref{eq:2.28}) appear.
The coordinates of these particles will be determined
by an equation similar to (\ref{eq:2.23})
\begin{equation}
   x_{p,l}(t) = x_{p,l}(t) + W_p( x_{p,l}) t ,  \label{eq:2.29a}
\end{equation}
with the velocity of particles determined by the function
$W_p(x)$ (see \ref{eq:2.27}), which is a sum of all larger modes,
and $ x_{p,l}$  are the coordinates of the zeros of the $p$th modes.
The difference between the coordinates of
adjacent particles $x_{p,l}$ and $x_{p,l}$
will then decrease with time like
$(1-t/ \mathaccent126t_{p+1})$,
where $\mathaccent126t_{p+1} =  {t_{p+1} ( 1 - \epsilon^{(n+3)/2}})$
is the inverse of the gradient of the function $W_p(x)$, see
(\ref{eq:2.27}) and (\ref{eq:2.28}).
Thus, the time of particle collision
for this generation will be described by equation (\ref{eq:2.29})
with $ p = p+1$, and the new masses will be determined by the period
of the $(p+1)$th mode - see equation (\ref{eq:2.25}).

The extrapolation of this particle merging process
to the next generations is evident by recurrence.
The $q$th collision of heavy particles takes place at time
$\mathaccent126t_{p+q} = t_{p+q} ( 1-\epsilon^{n+3 \over 2}$),
\ the masses of these particles at this time are determined
by the period of the $(p+q)$th mode \
$ m_{p+q} = \rho_0 L_{p+q} = 2 \pi \rho_0 / k_{p+q}$
(\ref{eq:2.8}), (\ref{eq:2.25}).
In the time interval
$t \in [\mathaccent126t_{p+q}, \mathaccent126t_{p+q+1}]$,
\begin{equation}
   {\mathaccent126t_{p+q+1} \over {\mathaccent126t_{p+q}}}
      =
    \epsilon^{-{(n+3)\over 2}} = N^{{n+3}\over 2};   \label{eq:2.30}
\end{equation}
the coordinates of  particles will be determined
by the equation (\ref{eq:2.29a}) with $p=p+q$. Here
$W_{p+q}(x)$ is the sum of the velocities of all
larger modes with $r>p+q$, and $x_{p+q,l}$ are
the zeros of $(p+q)$-th mode.
It is important to note that at time $t>\mathaccent126t _{p+q}$
the evolution of the particles is solely determined
by the modes with $r \geq p+q$.
It means, that at times $t> \mathaccent126t_{p+q}$
the position of the particles does not
depend on the presence in the initial condition
of the small scale modes with $ r < p+q$.

Thus, two processes with different initial velocities:
$ \mathaccent126 v_0(x)$, the field with small scales, and
$ v_0(x) $, the field without small scales modes:
\begin{equation}
   v_0(x) = W_{p+q-1}(x); \ \ \mathaccent126v_0(x) = W_{p-1}(x)
   \label{eq:3.30}
\end{equation}
will have the same evolution after $t>\mathaccent126t_{p+q-1}$.
Even if $p \rightarrow - \infty$
(when modes with very small scales $L_p \sim \epsilon^{-p} = N^p$
and very large amplitudes
$ a_p \sim a_0( \epsilon^{(n+1)/2})^p = a_0(N^{(n+1)/2})^{-p} )$
are present in the initial perturbation)
the multiple merging of the particles
will lead to the independence of the evolution of
large scale modes with respect to the small scale modes.

This effect is similar to the self--preservation
of large scale structures in Burgers turbulence
\cite{AGS93},\cite{GP99}.
When the initial field $v_0(x)$ is noise, the highly
nonlinear structures continuously interact and due to the merging
of shocks, their characteristic scale $L(t)$ constantly increases.
The presence of small scale noise perturbation $v_h(x)$ results in
additional fluctuations in the shock coordinates $\Delta x_k(t)$,
and these
fluctuations increase in strength with the passage of time.
Thus, the final result of the evolution of the field is
determined by the competition of two factors, the increase in the
external scale $L(t)$ of the structures and the increase in the
strength $\Delta x_k(t)$ of shock coordinates fluctuations,
the later being related to the perturbation $v_h(x)$.
In  a turbulence, having power index $n<1$ (\ref {eq:no4}),
multiple merging of shocks leads to
self-preservation of the large scale structures
independently of the presence of small scale components.
For the model signal this effect appears for arbitrary $n$
due to the special choice of wavenumbers and phases
of interacting modes.

It was stressed in the introduction that the solution of Burgers'
equation has a one-to-one correspondence with the dynamics
of the gas of inelastically interacting particles (\cite{GMS91}).
The stage when all light particles collide, forming heavy
particles, corresponds to the solution of Burgers' equation
with a well-defined slope $\partial_x v = 1/t$. In this case
the profile of the field $v(x,t)$ is fully determined
by the coordinates and amplitudes of the shocks.
Their coordinates $X_s(t)$ are
equal to the coordinates of heavy particles, their velocity
\begin{equation}
   v_s(t) = { d X_s(t)\over d t } = ( v_s(x_s-0,t) + v_s (x_s+0,t) ) /2
   \label{eq:2.26a}
\end{equation}
is equal to the velocity of the particles, and the amplitude of the shock
\begin{equation}
   \Delta v_s(x) = ( v(x_s-0,t) - v(x_s+0,t)) = m / t   \label{eq:2.27a}
\end{equation}
is determined by the mass of the particle
$(\rho_0 \equiv 1) $ (see, e.g., \cite{GMS91}).

Thus, the investigation of the motion of heavy particles permits
to fully reconstruct the properties of the velocity field $v(x,t)$
of Burgers' equation.

%=======================================================================
\section{The sawtooth Weierstrass-Mandelbrot
fractal function}\label{} \vskip 4mm

It was shown in the previous section that the evolution
of the particles (shocks) is determined
by the function $W_p(x)$  (\ref{eq:2.27}).
The basis functions of $W_p(x)$ are the reverse sawtooth periodic
functions with wavenumbers $k_r=k_0 \epsilon^r$ and amplitudes
$ a_r = a_0 \epsilon^{-hr}, h = -(n+1) / 2 $
satisfy relations (\ref{kpapa0}).
Wavenumbers form a geometrical progression
like in the Weierstrass function (see \cite{BerryLewis1980})
and accumulate at the origin $k=0$.
In the original Weierstrass function,
the situation was the opposite
with increasing frequencies, but nevertheless the function $W_p(x)$
has many properties of Weierstrass function
and of its generalisation -- the Weierstrass-Mandelbrot function
(see \cite{Mandelbrot77}, \cite{BerryLewis1980}).

We consider here a deterministic function $W_p(x)$
with the special phase relation $\phi_p= ( 2 \pi k / N ) p$ \ \ \
$(k=1,2,...,N; N=1/\epsilon$ ), thus, the discontinuities
in the largest modes $r>p+1$ coincide with some of the discontinuities
of the smaller mode $r=p+1$. The function $W_p(x)$ is continuous
in the intervals $2\pi/k_{p+1}=2\pi/(k_0\epsilon^{p+1})$
with the same slope in each interval. The inverse value of this
slope $\mathaccent126t_{p+1}$
\begin{equation}
   \mathaccent126t_{p+1}=t_{p+1}(1-\epsilon^{ {n+3}\over 2}); \ \
   t_{p+1}=t_0(\epsilon^{{n+3}\over 2})^{p+1}   \label{eq:3.1}
\end{equation}
is proportional to the nonlinear time $t_{p+1}$ of the
smallest mode. Of course, we need $n>-3$, so that the convergence of
(\ref{eq:2.28}) is assured and the inequalities $t_{p+1}>t_p$ hold.
The amplitudes of the modes are proportional to
$\epsilon^{(n+1)/2}$ and for $n>-1$ the function $W_p(x)$
is bounded
\begin{equation}
 W_p(x)\leq\sum_{r=p+1}^{\infty} a_r =
 a_0(\epsilon^{{n+1}\over 2})^{p+1}\ {1\over{1-\epsilon^{{n+1}\over 2}}}\ \
.
 \label{eq:3.2}
\end{equation}

Thus, for finite $p$ the energy of $W_p(x)$ is also
finite. For the  case of the phase relation introduced
above, the functions $W_p(x)$ also have scaling properties,
so that for instance for $k=0$, we have
\begin{equation}
   W_p(x)=\epsilon^{-hp} W_0 ( \epsilon^p x ); \ \
   W_p(\epsilon^mx)=\epsilon^{hm} W_{m+p} ( x ).   \label{eq:3.3}
\end{equation}

The case $ -1 < n < 1 $ is similar to the initial
conditions with generalized white noise in Burgers turbulence.
The energy of the initial signal in such turbulence
is determined by the largest cutoff wavenumber, so in our model
by the smallest scale $p$. If $ p \rightarrow -\infty$ the energy
of the model signal (as the energy of white noise) will tend to
infinity.  But from the considerations in the previous section we
have that at the finite time $t$ all the modes with $t_p<t$ have
finite energy $ \sim L_p^2/t^2$ due to the  nonlinear dissipation,
so that the whole energy of the turbulence is also finite.  Thus,
even in the case of "divergent" initial conditions $(p \rightarrow
-\infty )$, we will have a "convergent" solution for any time $ t > 0
$.

The case $n<-1$ is similar to having fractional Brownian motion
initial condition in Burgers turbulence.
In this case, the series (\ref{eq:3.2})  diverges and the initial signal
$W_p(x)$ is unbounded. But for Burgers turbulence
(~for the process of particles  motion and collisions~)
only relative velocity of the particles matters.
So we can use the same regularisation procedure
with $W_p(x)$. Such a procedure was done with
the Weierstrass function in \cite{Mandelbrot77}.

In our case, taking $\phi_p \equiv 0 $, we can introduce
the function $W_p^{\infty}(x) = W_p(x) - W_p(0)$,
according to \cite{Mandelbrot77},
which is finite in all finite spatial intervals.
The other way to get a bounded function is to use
special phase relations for the modes.

%=========================================================================
\section{Self-similarity properties of deterministic model
in one and two dimension}\label{}
\vskip 4mm

Here we summarise the properties of the evolution of the multiscale
deterministic signal using some additional information about
scaling characteristics of $W_p(x)$, and compare them with
the properties of the Burgers turbulence.

Let us consider the evolution of the multiscale signal
\begin{equation}
   v_0(x) = v_p(x) + W_p(x).   \label{eq:3.3a}
\end{equation}
It was shown that at times $t$ for which $t>\mathaccent126t_p$,
heavy particles with mass $ M_p = \rho_0 L_p$ appear and
their coordinates are determined by the relation (\ref{eq:2.30}).
These particles collide at time $\mathaccent126t_{p+1}$,  \ \
($\mathaccent126t_{p+1}/\mathaccent126t_{p} = N^{{{(n+3)}\over 2}},
N=\epsilon^{-1} $),
and new particles with masses $ m_{p+1} = \rho_0L_{p+1}=m_p N$
appear. Their motion will be determined by the
same law (\ref{eq:2.30}) with substitution $ p \rightarrow p+1$.
Using the scaling properties of $W_p(x)$   (\ref{eq:3.3})
we have, that the motion of the particles in this interval
will be similar to the motion of the particles
in the interval $  [ \mathaccent126t_p, \mathaccent126t_{p+1}]$\
if we rescale the time
$ t / \mathaccent126t_p \Rightarrow t / \mathaccent126t_{p+1} $.
Since the ratio $t_{p+1}/t_p$ does not depend on $p$, one
can speak about the logarithmic periodic self-similarity of
the motion of the particles.
This means that at
arbitrary interval $ [ \mathaccent126t_q, \mathaccent126t_{q+1} ] $
the motion of the particles will be similar
to the motion of the particles in the interval
$ \mathaccent126t_p, \mathaccent126t_{p+1}$, by the scaling factor
$x_p/x_q = \epsilon^{p-q}$ in space, and  the scaling function
$t_p/t_q = (\epsilon^{ - { {n+3} \over 2}   })^{p-q}$
in time. The coordinates and masses  of the
particles  fully determine the velocity field, and so
the solution of Burgers' equation is also
logarithmic periodic self--similar.

With each collision, the mass $M(t)$ of the particles increases
$N=1/\epsilon$ times. The time interval between the
two successive collisions increases as $t_{p+1}/t_p=N^{{n+3}\over 2}$.
Thus, by the approximation of piecewise constant function $m(t)$
by the power law
\begin{equation}
   m(t) \simeq m_0(t/t_0)^{(n+3)/ 2 }.   \label{eq:mas}
\end{equation}
we obtain the same result as for the Burgers turbulence.
In our case, $m$ is proportional to the period
of the smallest mode at time $t$, and is analogous to
the integral scale in Burgers turbulence.

In the case $n>-1$ we can also estimate
the energy decay of the model signal. For $n>-1$
and $\epsilon \ll 1$ the main energy of the signal at time $t$
is in the smallest mode and is proportional to $L^2 (t) / t^2$.\
Thus, we have here again the same law for the energy decay as
for Burgers turbulence.

The numerical simulation based on the algorithm
\cite{TroussovLinAlg96} was done to illustrate the process
of particles merging and velocity field evolution.
The trajectories of the particles and profile of the field
at different times are plotted
for the initial "white noise" signal  ($n=0,\ h=-1/2$) in Fig.3,
and for the initial "Brownian" motion ($n=-2,\ h=1/2$) in Fig.4.
Ten modes with the ratio of successive wavenumbers
$\epsilon = 1/N = 1/2$ were used.
The plots show the initial stage of the evolution
in some relatively small region where the finiteness of
the number of modes is not significant.

In Fig.~3 one can see that for $n=0$ the initial
"sawtooth" multiscale function oscillates near $v=0$
like a "white
noise" with finite variance.  After the collision of light
particles, when the reverse sawtooth function transforms into
a sawtooth wave with positive gradient $\partial_x v = 1/t$, the
structure of the signal is relatively simple, and even for $N=2$ the
main energy remains in the mode with smallest wavenumber.

In the case $n=-2$ the initial profile has a large deviation
behaviour which
is typical for Brownian motion functions.
After merging of light particles, the sawtooth profile has a set of
small shocks with different amplitudes, which is also similar to the
properties of Brownian signal in the Burgers turbulence \cite{VeDuFrNo94}.

In Figs. 3(c) and 4(c) the velocity field at three
successive merger times
$t_* / t_{**} = N^{(n+3)/2}$ are plotted. These figures
show the logarithmic periodic self-similarity of the evolution
of multiscale signals.

We  notice  now that such multiscale waves may be constructed
for multidimensional Burgers' equation.
Let us assume that the initial vector
field ${\bf V}_p \left( {{\bf x}} \right)$
is  an infinite series of  "reverse" modes
${\bf v}_r \left( {{\bf x}}\right)$:
\begin{equation}
    {\bf V}_p \left( {{\bf x}} \right)
    =\sum_{r=p}^{\infty}{\bf v}_r \left( {{\bf x}} \right),
     \label{eq:Wvec}
\end{equation}
In the two dimensional case,
the $r$th "reverse" mode  may be composed
of piecewise linear functions defined on a system of regular
triangles of size $L_r$ covering the plane.
We consider here the special
case when the ratio of the scales
of two adjacent modes is $L_{r+1}/L_r  =\epsilon^{-1}=N = 2 $.
We assume also the special symmetry and phase relation between the
different modes. In our case one big triangle is divided into
four smaller triangles
with vertices located at midpoints of its sides (~see Fig.5~).
We assume that inside each triangle in the $r$th mode the
velocity has a linear profile
$ {\bf v}_r \left( {{\bf x}} \right) =
 - \gamma_r \left( {\bf x} - {\bf x}_ +  \right)$,
where ${\bf x}_ + $ is the coordinate of the center of the
triangle.
The solution of the multidimensional Burgers equation  for such
initial perturbation
\begin{equation}
{\bf v}\left( {{\bf x},t} \right) =
\frac{{ - \gamma \left( {{\bf x} - {\bf x}_ +  } \right)}}{{1 - \gamma t}}
 \label{eq:vecv}
\end{equation}
is now valid inside the triangle of size
$L_r(t)=L_r (1-{t\gamma}_r)$ .
The value $\gamma_r^{-1}$ has the dimension of time, and
at the finite time $t_r = \gamma_r^{-1}$ the velocity gradient
becomes infinite.

On the other hand, at time $t = t_r$,
the mode transforms into a "direct" sawtooth wave with
the universal behaviour inside the new set of
triangles and
with the gradient ${1/t}$
independent of the amplitude and wavenumber of the mode:
\begin{equation}
{\bf v}\left( {{\bf x},t} \right) =
\frac{{\bf x} - {\bf x}_ c}{t},
 \label{eq:vecvl}
\end{equation}
where  ${\bf x}_ c$ is now the center of the triangle,
coinciding with the top of the initial triangular set.
Consider now the evolution of a gas of sticky particles
in the case of independent evolution of the $r$th mode.
Then, it is obvious that at the initial stage, inside the
"collapsing" triangle of size $L_r(t)=L_r (1-{t\gamma}_r)$
the density increases as
\begin{equation}
 \rho({\bf x},t) = \rho_0 { 1\over { (1 - {t/t_p})^2}},
     \label{eq:rho2}
\end{equation}
while $\rho$ is zero outside "collapsing" triangular in each cell.
At time $t=t_r$, all the light
particles in each cell collide into a single heavy particle with mass
\begin{equation}
   m_r = \rho_0 L_r^2\sqrt3/4 ,
     \label{eqm2}
\end{equation}
and the heavy particles'  positions  are equal to the center
of the initial
triangle ${\bf x}_ + $.

We assume also that the evolution of the $r$th mode is characterised
by a nonlinear time $t_r$ the same as in the one
dimension case (\ref{eq:2.5}):
\begin{equation}
    t_r = \gamma_r^{-1}  = t_0/( 2^{-(n+3)/2})^r.
    \label{eq:time}
\end{equation}

Let us now consider the evolution  of the vector
field ${\bf V}_p \left( {{\bf x}} \right)$ (\ref{eq:Wvec})
which is  an infinite series of  "reverse" modes.
The evolution of the vector field is very similar to
the evolution of the scalar field (\ref{eq:2.27}). For the gradient
of the initial perturbation ${\bf V}_p \left( {{\bf x}} \right)$
we have the same relation (\ref{eq:2.28}) as in  $1D$. The effective
time of nonlinearity of the smallest $p$th mode in  the
vector field (\ref{eq:Wvec}), in presence of all large scale
modes,
is determined  by the equation (\ref{eq:2.29}).
Thus, after the time of collision $\mathaccent126t_p$, heavy particles
with mass $ m_p $ (\ref{eqm2}) appear.
Velocities of these particles will be determined
by the function
${\bf V}_{p+1} \left( {{\bf x}} \right) (\ref{eq:Wvec})$,
which is a sum of all larger modes, but the number of
particle collisions
is determined by the next $p+1$ mode.
At time
$\mathaccent126t_{p+1}$ we have a collision of four
heavy particles.

The extrapolation of this particle merger process
to next generations is evident by recurrence.
The $q$th collision of heavy particles takes place at time
$\mathaccent126t_{p+q}$), (\ref{eq:time})
the masses of these particles at this time are determined
by the scale of the $(p+q)$th mode \
$ m_{p+q} = \rho_0 L_{p+q}^2\sqrt3/4$ (\ref{eqm2}).
Here also one
can speak  about the logarithmic periodic self-similarity of
the motion of the particles.
This means that at
arbitrary interval $ [ \mathaccent126t_q, \mathaccent126t_{q+1} ] $
the motion of the particles will be similar
to the motion of the particles in the interval
$ \mathaccent126t_p, \mathaccent126t_{p+1}$, by the scaling factor
$x_p/x_q = 2^{(q-p)}$ in space, and  the scaling function
$t_p/t_q = (2^{ { {n+3} \over 2}   })^{(p-q)}$
in time.

We used computer simulation for studying two dimensional
case; results of the simulation were generated in so called
VRML (Virtual Reality Modeling Language), which enables to
handle three dimensional figure in different projections. Fig. 6
presents snapshots of this modeling. On the Fig. 6(a) one
can see that particles formed by small triangles move towards
the center of an embracing triangle; this center in its turn,
moves towards the center of the next bigger triangle in the
hierarchy; e.t.c. Fig.6(b) gives the side-view of this process.

\section{Conclusion}\label{} \vskip 4mm

In conclusion, we would like to point out  that
the evolution of the multiscale signal
with the Weierstrass spectrum simulates
properties of Burgers turbulence such as self-similarity,
conservation of large scale structures
and has the same laws of the energy decay and integral scale.
The difference between the deterministic model and
Burgers turbulence is that here we have
the exact solution for the evolution of multiscale signals
and these properties are not stochastic but deterministic.
The evolution of the multiscale signal is exactly self-similar
in logarithmically spaced time intervals.
The evolution of the large scale modes is completely independent
of the small scales modes, even if these have very large amplitudes.

These properties take place for Burgers turbulence in the
stochastical
sense and, moreover, for a signal with cutoff frequencies
of small scales, only asymptotically.
Of course, these properties of the multiscale signal are
determined by
the special form of modes (reverse sawtooth function),
the special relations between wavenumbers of modes
($k_{r+1} = k_r / N$, where $N$ is an integer)
and their phase relations.

On the other hand, these model signals do not
reflect such properties of Burgers turbulence as qualitative
difference in the behaviour of the turbulence for
$n<1$ and $n>1$ in the power spectrum (\ref{eq:energy}) due
to the process of generation of large scale components in the spectrum.
For the deterministic model this process is not present
due to the special relation of wavenumbers.

Let as now move to the mechanical interpretation of solution
of the Burgers equation.
For the initial reverse sawtooth wave,
all the matter turns into heavy particles
at the same moment of time.
Thus, after this time, the evolution of Burgers turbulence is
fully determined by the motion of heavy particles.
The trajectories of heavy particles form regular tree-like structure
on the plane $(X,t)$, see Figs. 3 and 4.
The properties of this structure depend on the parameters of our model.
The integer $ \epsilon^{-1} = N$ is the number
of trajectories which intersect at one point
and form a new branch of our structure.
For $ \epsilon^{-1} = N = 2 $ we, thus, obtain binary tree structure.
Changing of the parameter $h$ stretches or contracts the structure
in the $t$ direction.
One can say that our structure is the plane representation
of the $N-$tree;
the root of our tree is located at $t=+\infty$.
This tree is similar to the flattened fractal model of
botanical umbrella tree (see \cite{Mandelbrot77}).
If we take some node $(X,T)$ of this structure
as a root, and consider the trajectories of all
heavy particles, which will merge
at the moment of time $T$ at this point $(X,T)$
we shall also obtain an $n$-tree.
The whole tree seems self-similar, because every branch plus the
branches it carries is a reduced scale version of the whole.

We also notice  that such multiscale waves may be constructed
for multidimensional Burgers' equation.

%=======================================================================
\section{Acknowledgements}\label{Ackn} \vskip 4mm
The authors are grateful to U.~Frisch
for useful discussions and for his hospitality at the
Observatoire de la C\^ote d'Azur,
to G.M.~Molchan, A.I.~Saichev, A.~Noullez and
W.A.~Woyczynski
for useful discussions.
This work was partially supported by the French Ministry of Higher
Education, by
INTAS through grant No 97--11134, by
RFBR through grant 99-02-18354.

%=======================================================================

%=======================================================================
\vfill\eject
% \section{Figure captions}\label{}
\noindent{\Large\bf Figure captions}   \vskip 4mm

\vskip 3mm\noindent
Figure 1:
Evolution of one-mode pulse. (a) particle trajectories;
(b) evolution of the initial velocity field
given in the same spatio-temporal scale.
Bold lines on the time axis denote moments
at which the profiles of the velocity are plotted.

\vskip 4mm\noindent
Figure 2:
Evolution and interaction of two modes.
(a) particle trajectories; (b) evolution of
the initial velocity field given in the same spatio-temporal scale.
Bold points on the time axis denote moments
at which the profiles of the velocity are plotted.

\vskip 4mm\noindent
Figure 3:
Evolution of the multiscale fractal signal with $n=0 \ (h=-1/2)$,
corresponding to "white noise" signal.
(a) particle trajectories;
(b) evolution of the initial velocity field;
(b) velocity field taken at the initial moment of time
and then at three successive time moments of self-similarity.

\vskip 4mm\noindent
Figure 4:
Evolution of the multiscale fractal signal with $n=-2 \ (h=1/2)$,
corresponding to "Brownian motion" signal.
(a) particle trajectories;
(b) evolution of the initial velocity field;
(b) velocity field taken at the initial moment of time
and then at three successive time moments of self-similarity.

\vskip 4mm\noindent
Figure 5:
Plane construction for two dimensional case. The
hierarchy of triangles, used for the construction of the multiscale
signal, with four layers shown. The initial signal is constructed
as a series of signals piece-wise linear on triangles.

\vskip 4mm\noindent
Figure 6:
Particle trajectories for the multiscale fractal signal in
two dimensional case presented in spatio-time three dimensional
space; the width of particle trajectory reflects its mass:
(a) top view;
(b) side-view.
\vfill\eject

\end{document}